# The model is the message: Lightweight convolutional autoencoders applied to noisy imaging data for planetary science and astrobiology

*Caleb Scharf, NASA Ames Research Center (caleb.a.scharf@nasa.gov)*

**Abstract:**

The application of convolutional autoencoder deep learning to imaging data for planetary science and astrobiological use is briefly reviewed and explored with a focus on the need to understand algorithmic rationale, process, and results when machine learning is utilized. Successful autoencoders train to build a model that captures the features of data in a dimensionally reduced form (the latent representation) that can then be used to recreate the original input. One application is the reconstruction of incomplete or noisy data. Here a baseline, lightweight convolutional autoencoder is used to examine the utility for planetary image reconstruction or inpainting in situations where there is destructive random noise (i.e., either luminance noise with zero returned data in some image pixels, or color noise with random additive levels across pixel channels). It is shown that, in certain use cases, multi-color image reconstruction can be usefully applied even with extensive random destructive noise with 90% areal coverage and higher. This capability is discussed in the context of intentional masking to reduce data bandwidth, or situations with low-illumination levels and other factors that obscure image data (e.g., sensor degradation or atmospheric conditions). It is further suggested that for some scientific use cases the model latent space and representations have more utility than large raw imaging datasets.

**1.0 Introduction:**

Imaging plays a central role in planetary science and exploration, and in astrobiology, where specific, sometimes novel, environmental features are sought as markers of geochemistry, habitability, or biogenicity. The use of imaging applies to remote-sensing orbital instruments, or in situ devices. However, large image datasets can lead to intensive onboard processing and storage demands, as well as larger downlink bandwidth requirements, all which impact mission and instrument design, particularly for lower-cost exploration approaches, or locations in the outer solar system. These data can also be subject to noise or gaps, and are highly dependent on local conditions (e.g., varying lighting levels or viewing angles, as well as atmospheric conditions, etc.). Additionally, for multi-spectral data, different camera sensors or filters may not always be available for simultaneous data capture or for complete coverage of a scene (e.g., as seen with Mars Reconnaissance Orbiter (MRO) CRISM data, Seelos et al. 2024).

Challenges with imaging data are likely to grow as the future exploration of planetary bodies in the solar system produces increasingly large datasets of all types and makes use of increasingly autonomous and diverse robotic orbiters, landers, and mobile exploration platforms. In many existing cases this already calls for prioritization of what data is gathered in-situ, and what data is returned to Earth. Computer-aided vision and image processing have a long history, and advances in machine-learning (particularly deep learning approaches) have added greatly to the range and flexibility of near real time image processing and analytical methods that are now available [Chai et al. 2021]. For example, the Autonomous Exploration for Gathering Increased Science system (AEGIS), on the Perseverance and Curiosity Mars rovers, parses much of the navigational imagery [Francis et al. 2017] and, in effect, performs triage on activities to increase efficiency in deploying scientific instruments and making targeting decisions for its instruments. This can also reduce the throughput rate needed for data exchange with remote controllers, helping unburden orbital relays and deep space communications.

At the same time, the use of machine learning to support in situ scientific analyses, or downstream analyses for planetary and astrobiological research, is still an emerging area, without many, if any, clear data standards or standard models and approaches (see references in Sections 1.1, 1.2 below, also Scharf et al. 2024). In part, this is due to a lack of broad community experience with the specialized knowledge necessary for developing applications that use any of the numerous different machine-learning approaches that exist. It is also due to the constantly evolving landscape of these approaches. There is also a fundamental, qualitative difference between applying traditional tools of analysis and more elaborate machine learning tools to any data. The former can be applied with fully standardized measures of success, and comparatively few 'tuned' parameters (e.g., simple minimization of variances in function fitting), while the latter can involve many rounds of experimentation to build a successful and useful model, and to optimize its performance (with many hyperparameters to adjust) for a particular class of problem.

In this context, given that there is a wide range in the level of experience with machine learning approaches across the scientific community, Section 1 and subsections below present a brief review of imaging data challenges and machine learning approaches relevant to planetary science and astrobiology, and to the work presented here.

## 1.1 Intelligent compression and reconstruction from noisy data

Traditional algorithmic compression can reduce data volumes. However, these approaches introduce further challenges for scientific applications. Industry-standard data compression algorithms, such as the lossless JPEG2000 standard, or the Lempel–Ziv–Welch (LZW) algorithm (used in the GIF image format) typically offer lossless compression ratios of 2:1 for moderately high-entropy content imagery. These algorithms work agnostically on image data without broader contextual information about the nature of features in the data (where a feature is a measurable property, variable, or attribute of the data), or about what the downstream use of data will be. That lack of feature context becomes more problematic with increasingly aggressive, lossy compression that is, in effect, a mixture of triage and compression, where informational features in the data become more and more degraded. Traditional compression can also be problematic in the presence of noise or incompleteness that may be treated as features by the compression algorithms.

Traditional approaches to handling noise range from filtering (e.g. high-frequency spatial filters), to interpolation, or inpainting if the noise is destructive and results in empty image regions [for a review see Jam et al., 2021]. A more sophisticated, mathematically rigorous, and optimal approach involves maximum entropy methods (going back to the 1970s and 1980s, e.g., Frieden 1972) for reconstructing sparsely occupied fields of view, or data that requires deconvolution of a blurring function, and can often achieve super resolution effects. A Shannon entropy is summed over pixels in an image to yield the most mathematically robust way to select a single (permitted) image from a vast number that might fit the available data – i.e., the image that has maximum entropy or is maximally "honest" about the uncertainty of the reconstruction. This becomes a constrained non-linear optimization problem in a vector space [Skilling & Bryan 1984]. However, maximum entropy can be computationally demanding and makes assumptions about the underlying probability distributions of pixel values and noise (e.g., Gaussian or other distributions).

By contrast, in broad terms, machine learning approaches – such as deep-learning models - can learn to build an agnostic internal representation, or multi-dimensional space, matching the statistical, contextual, informational features in data (typically a corpus of data). That representation, or space, can be encoded to reduce noise or gaps in subsequently presented data. A classic training application is cleaning up noisy images of written characters, or to train for classification tasks [Lecun et al. 1998] (see also 1.2.2 below). Therefore, deep-learning methods have similarities with, but not exact correspondence to, maximum entropy approaches, in the sense that both capture information about the features of the data corpus and utilize this to optimize their reconstruction of degraded inputs. A deep learning loss function can also be based on entropy [e.g., Baggenstoss 2021] with reductions in mean square reconstruction error up to a factor of two over other approaches.

In this present work a deep learning approach is used to build feature representations of planetary imaging data and train a baseline, lightweight model that can use its encoded feature space to reconstruct new data that is subject to different types of strong destructive noise (Section 2 below). The capabilities of this model are examined as an example of how machine learning approaches can support planetary and astrobiological exploration. This is further examined as a guide to potential non-traditional approaches to data acquisition where it is the informational features in data that matter and, consequently, in certain situations the model representation of those features could replace a large corpus of raw imaging data.

## 1.2 Convolutional autoencoders and masking

### 1.2.1 Review of autoencoders and planetary science applications

One of the fundamental architectural forms of unsupervised machine learning models is an autoencoder that deploys a layered, connected perceptron network scheme as an *encoder* to perform learning of features in unlabeled data [e.g., LeCun, 1987; Gallinari et al., 1987]. A successful encoder model maps the abstract, hidden, or "latent", features in data as multi-dimensional vectors (a "latent representation") projected into a so-called "latent space"; an embedding within a high-dimensional manifold that nonetheless has lower dimensionality than the feature space of the data used to train on. The multi-dimensional latent representation attempts to capture all features in data in a way that preserves the most information-rich relationships, enabling tasks like classification or prediction. In other words, a latent representation serves as a compressed, or summarized, form of the input data, that contains only the necessary features for the model to perform its task. The learnt, abstract latent space (also sometimes, confusingly, referred to as the latent representation), is held in the weights and biases within a model and is the result of a model's attempt to characterize the underlying structure or patterns in the data corpus.

An autoencoder model expands on this approach via a further, layered *decoder* scheme that often mirrors the encoder in its structure. The decoder attempts to reconstruct instances of new input by using only its latent representation, and the entire autoencoder learns the best latent space for this goal. Autoencoder models therefore employ a mild form of generative learning, in the sense that they iterate to minimize the difference between input and generated output by adjusting their internal parameters and their encoded latent space for this task.

For data with complex, multidimensional features, such as multispectral imaging data, a *convolutional* autoencoder (CAE) is a common, relatively straightforward model to deploy (see for example the seminal paper by Zeiler and Fergus (2013) on how CAEs work and why they work so well). In a CAE for image data, encoder layers operate by convolving or scanning learnable spatial filters (or kernels) across each image channel (e.g., RGB color channels or multispectral channels). The convolution (a summed product of image and filter pixel values) is recorded at each pixel location. This new image array is then typically reduced in size for subsequent layers by pooling the pixel outputs of the convolution filters after passing through intermediate "activation" layers that use a non-linear perceptron response function such as a rectilinear or sigmoid response. That non-linear response helps the CAE to handle noise and complex features in the data. This dimensionally reduced array can then be passed on to further convolution, activation, and pooling layers, each further reducing the spatial dimensions of the data to eventually produce the latent representation. The number of such layers and the number of filters used in each layer are chosen according to need and are some of the hyperparameters of the CAE. For example, the number of filters may increase in each layer to better capture increasingly complex, higher-order features in the data.

The decoder layers use further convolutions or sometimes transpose convolution (which is not the same as the usual mathematical definition of deconvolution, Dumoulin & Visin, 2016) and "up sampling" to restore the original image size (e.g., by padding a higher resolution layer with repeated neighboring pixel values). The difference between that decoded output and the original input yields errors that can be backpropagated through the CAE layers to attempt to improve the reconstruction. This training procedure can use a variety of techniques to optimize the CAE (see implementation below for an example).

Consequently, in all CAE encoder and decoder layers the filter weights are learnable, along with the learnable perceptron bias values in the activation layer, and together these train to encode the latent space of all the training data. Filters are often in the form of a 3 x 3-pixel array of weights. These weights are therefore analogous to the weights in a connected multi-layer neural network, but in the CAE the effective inter-layer connections have spatial/neighbor constraints and a complicated and distributed topology across the whole architecture.

The power of autoencoders and CAEs lies in their flexibility of use for a wide range of problems that can utilize the modeling of features. For example, image classification models can train on the multidimensional vector formed as an input image is passed through the encoder since that vector represents a distillation of informational features. Autoencoders and CAEs can also be applied to more complicated and useful generative problems. For example, a CAE can be trained on artificially degraded inputs with a target output of the undegraded data and therefore learn to generate the best reconstruction of the undegraded data (see Section 1.2.2. below). Somewhat counterintuitively this can improve the robustness of the learning, since success comes from finding the most important (informationally rich) features in the data to fill in where data is missing or corrupted (see also 2.2. below).

A similar application is as a generative tool to predict complexly-correlated properties from data subsets or channels. In planetary science, for example, CAEs have been used with additional learning components to predict velocities from seismic data [Chen, Saygin 2021]. Other uses include anomaly or novelty detection, where the decoder output – that can be considered as a statistically most probable reconstruction - deviates strongly from the input data. Using autoencoders for novelty or anomaly detection in planetary imaging data has been explored in a variety of previous works, see for example Stefanuk and Skonieczny (2022), Kerner et al. (2019a, b), and specifically for Mars rover instrument targeting [Wagstaff et al 2020]. Applications also exist for abundance mapping of minerals inferred from hyperspectral planetary surface data [Hipperson et al. 2020].

Other related deep learning applications in planetary science have included the application of a CAE to landform classification and detection of planetary skylights [Nodjoumi et al. 2022]. A generative model named HORUS has been used to de-noise and predict image content from Lunar Reconnaissance Orbiter data of transiently shadowed regions or permanently shadowed regions (PSRs) on the lunar surface, where existing data consists of photon detection at the limits of instrument sensitivity [Bickell et al. 2021]. Transfer learning is another key use case for autoencoders. For example, CAEs have been applied to enhance and denoise Titan SAR data following training on equivalent Earth Landsat terrain data (e.g., dunes). Although this clearly introduces (known) biases, it also makes use of the presumed universality of certain landform physics [Holland et al. 2022].

A further elaboration of any autoencoder is a variational autoencoder, as a type of variational Bayesian method (see Discussion section 3.0 below). For image inputs this entails each pixel being mapped through the layers to build a *distribution* in a persistent latent space rather than to a single point in that space for a given input. The latent representation that is formed is therefore a multi-dimensional probability distribution of features. This has use for efficient simulations or test data for other analysis methods and has been applied to novelty detection in planetary image data [Stefanuk and Skonieczny, 2022].

### 1.2.2 Masked autoencoders

A powerful training strategy for autoencoders and other deep learning models is to deploy deliberate masking of input data (partial removal or blocking of some information), or to add noise to input data. This forces the autoencoder to learn more robust and meaningful representations to improve its success at inferring what exists in missing or obscured parts and to reconstruct a 'clean' version of the input. Consequently, a model can become better at generalizing from the features in training data. Masking in image modeling (and other data types, particularly language modeling, Devlin et al 2018) has been shown to outperform earlier self-supervised methods (see e.g., Gao et al 2022, He et al. 2022, Xie et al. 2022). Furthermore, a variety of studies have shown that masked autoencoders perform particularly well on images (e.g., Huang et al. 2024, He et al. 2022) and videos (e.g., Feichtenhofer et al. 2022, Huang et al. 2023). Not only is masking a valuable strategy for model architectures that include the CAEs discussed here, it appears to work well with more recent approaches such as diffusion-based modeling (e.g. Wei et al 2022) and models that employ attention-based transformer techniques (e.g., Hou et al., 2022; Bachmann et al., 2022).

In this present work, the utility of a baseline, lightweight masked CAE is explored for reconstructing common planetary imaging data subject to destructive noise that could be a consequence of instrument behavior, deliberate sparse sampling/masking, reduced scene illumination or other optical effects of an environment. Following from this, the potential for CAEs to reduce data bandwidth requirements is discussed. Options are then considered for future scientific exploration enabled by modifying the ways that data is utilized; where the feature contents are of primary importance and latent representations could replace traditional forms of data reduction and storage.

It should also be emphasized that a baseline CAE is far from the current state-of-the-art in image reconstruction, where leading models include U-Nets (building on CAEs, with an emphasis on image segmentation, see Ronneberger et al. 2015) and Transformers or diffusion models (capturing larger scale dependencies, and learning through gradual noise removal, respectively). However, as discussed above and in the following sections, the availability of planetary training data, the 'lightweight', comparatively simple nature of CAEs, and the storage and bandwidth requirements that can exist for exploration, indicate that performance gains from larger, more complex models may not always outweigh other factors in the trade space of mission design.

## 2. Methods

### 2.1 A baseline, lightweight CAE model

To balance computational demands against data volume, the CAE built here takes as input 128 x 128 pixel, 3-channel (e.g., RGB) images. Four encoder layers each consist of filter layers, corresponding activation layers using the Rectified Linear Unit (ReLU) activation function, and a 2 x 2 Max Pooling layer with 2 x 2 stride (choosing the maximum pixel value in each 2 x 2 region of the activation layer) to reduce spatial dimensions through each layer. Filter numbers in the encoder layers are 16, 32, 64, and 128, following the standard practice of increasing filter number as higher-order feature relationships are encoded deeper in the network. Filters have spatial dimensions of 3 x 3. The decoder mirrors the encoder with 4 layers and filter numbers in sequence as 128, 64, 32, 16 using upsampling with 2 x 2 upscaling to finally output 128 x 128, 3-channel reconstructed images. The code utilizes the common libraries: cv2, tensorflow, and keras (see Github project page for associated code, https://github.com/caleb-nasa/gt-CAE).

The training regime uses 5,000 subimages selected randomly from a master training image (i.e., a single selected high-resolution image, see Figure 1 below), with overlaps allowed. This training set is augmented using the ImageDataGenerator preprocessing function that includes modest (<20%) rotation, width/height modification, shear, and horizontal flipping to improve the model's ability to generalize. The CAE is initially trained using batch sizes of 32 over a total of 40 epochs, with the Adam optimization (a commonly used alternative to stochastic gradient descent, that adjusts learning rates for each parameter individually, Kingma and Ba, 2014), and a mean squared error loss function (the measure of difference between input and output). Once the training and validation losses over 40 epochs are tabulated the final number of training epochs used is adjusted to optimize the loss behavior (see below).

As built, this CAE is relatively basic and does not employ many of the more sophisticated algorithmic approaches that are available for tasks like pooling, upscaling, or activation. Other hyperparameters are also set to relatively generic levels without further inspection. The goal of this present work is to explore generic, baseline properties of a CAE in the context of planetary imaging and astrobiology.

### 2.1.1 Fitting and testing

To set a baseline for the CAE behavior an initial training was performed on non-noisy inputs (i.e., the CAE is simply trained to reproduce the input). A representative image/scene from the Perseverance rover's Mastcam on Mars (Figure 1) was chosen as an example of a complex, but relatively uniform terrain under non-extreme lighting conditions (i.e., the Sun was at moderate elevation). As noted previously, this present work is not intended to develop a generalized model for Mars's highly diverse surface scenes, but to instead examine a proof-of-principle. The master image was divided into a training image and a smaller validation/unseen image; each were further subsampled for training and validation as described above.

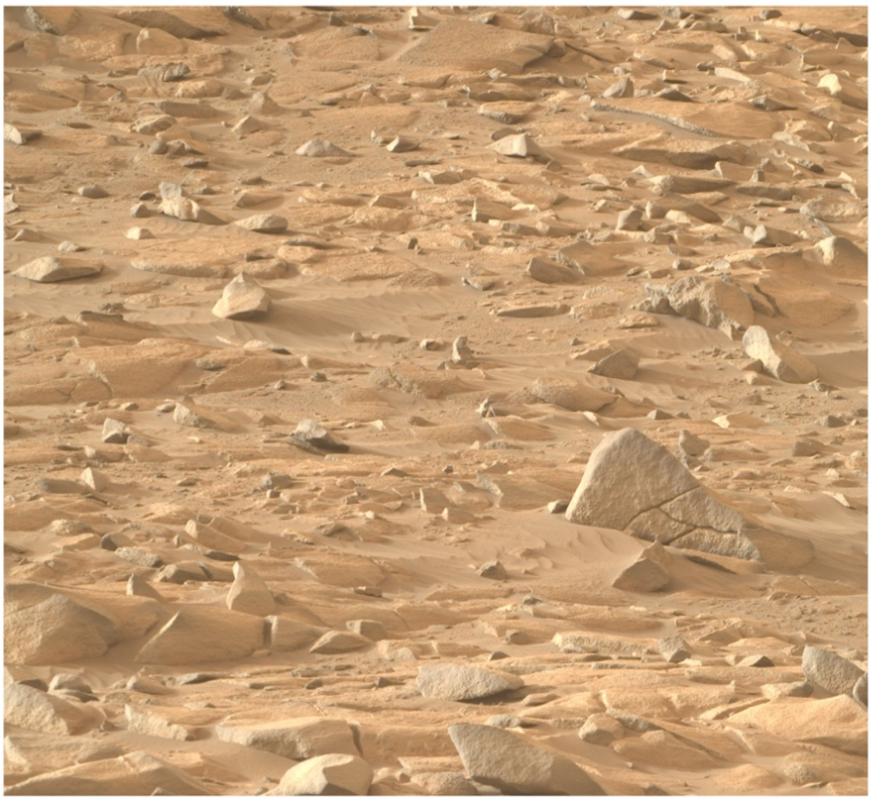
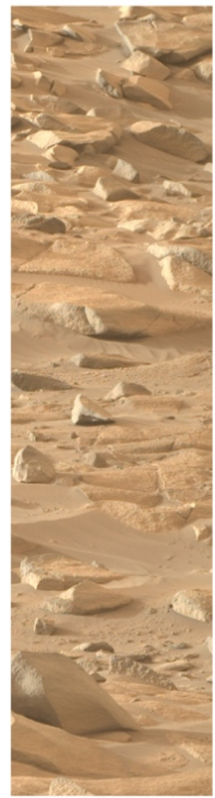

Training master    Validation master

**Figure 1**: *Mars Perseverance Mastcam mage data used in this paper. Left panel shows the training master from which subimages are drawn for CAE training (1282 by 1173 pixels). Right panel shows the validation or unseen image segment that is then further subsampled (289 by 1173 pixels). Image identifiers: Sol 1159, May 24, 2024, LMST 09:14:47 Right Mastcam-Z Camera, Image source:* [https://mars.nasa.gov/mars2020/multimedia/raw-images/ZR0_1159_0769818046_473EBY_N0523018ZCAM09196_0630LMJ](https://mars.nasa.gov/mars2020/multimedia/raw-images/ZR0_1159_0769818046_473EBY_N0523018ZCAM09196_0630LMJ) *(NASA/JPL-Caltech/ASU).*

A validation set consisting of 16 unique subimages from the unseen part of the master image was used to provide validation loss estimates via the history data during the autoencoder training. Figure 2 plots the training losses and validation losses as a function of training epoch. Both show acceptable behavior, with the final losses very close in value at the end epoch, differing by < 25%, indicating that there is no significant over or under fitting to the data. Since the validation set is small there is some noise in the validation loss curve (e.g., at epoch 8 where there appears to be overfitting). Further training might continue to improve training losses but will likely increase validation loss more than is desirable and therefore the model will be less able to generalize. Based on these loss functions, this basic test of the CAE demonstrates satisfactory behavior after ~10 training epochs.

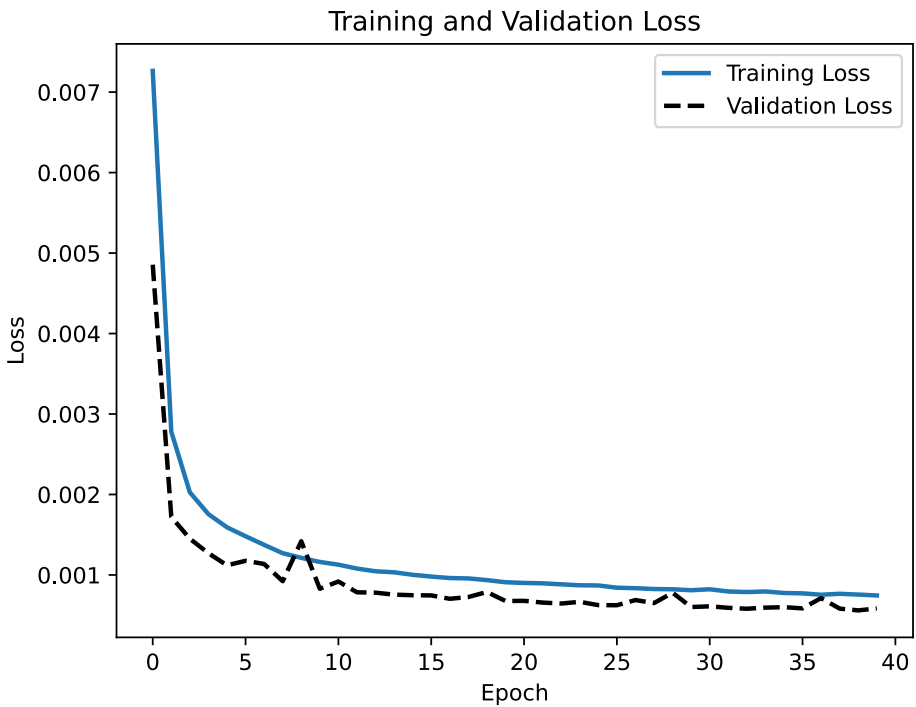

**Figure 2**: *Training and validation losses for the test CAE (with image inputs equal to image outputs) plotted versus training epoch (40 total).*

## 2.2 Reconstruction with destructive luminance noise or sparse sampling

As outlined in Section 1, destructive luminance noise (e.g., image regions with zero data) from instrument limitations or transmission problems, or constraints due to low light conditions or atmospheric conditions, are a potential challenge to traditional imaging and sensing analyses in planetary science. Similarly, deliberate sparse-sampling or masking of image data could be a useful alternate methodology for reducing storage and bandwidth needs in certain cases, e.g., quick-fire scene evaluation.

Traditional techniques of "in-painting" gaps in images have often been limited to simple extrapolation or interpolation using running averages or function fitting [Jam et al. 2021] and can struggle for cases of globally destructive noise where adjacent image regions are themselves highly incomplete. CAEs trained to reconstruct images from noisy/degraded input have long been a standard application in data science. Furthermore, unlike the example of simple recovery of an original input (2.1.1 above), training to recover incomplete or adjacent data (including correlated spectral bands) produces more generalizable models, as described in 1.2.2 above.

### 2.2.1 Training on fixed luminance noise levels

The CAE was trained on the same Mastcam master image to generate 6 models; each learning to reconstruct inputs with randomly 'destroyed' or masked (zeroed out) pixels with a fixed total areal noise coverage of, respectively, 50%, 75%, 90%, 95%, 98%, and 99% of the original images. These were all trained using 10 epochs (see discussion above) as a conservative choice to reduce compute time and to avoid overfitting.

Figure 3 presents example results for 5 of the models using a previously unseen subimage drawn from the validation master image (the 95% level model is omitted for clarity of presentation). The CAE succeeds at recovering gross features even at 99% masking of the input data, and at up to 90% masking, the reconstructed images are qualitatively very similar to those with lower noise levels. Error maps formed from differencing the reconstructions from the original images (Figure 3) show that as noise levels increase, the CAE fails to accurately recover larger scale features, as would be expected since the masking obscures increasingly large-scale spatial information.

To attempt to better quantify the CAE efficacy a simple metric is the mean over all image pixels of the absolute difference (i.e. a positive number) across each of the RGB channels between the original and reconstructed image, where pixel channel values are in the range [0,1]. This is plotted in Figure 4a as a function of the noise/masking level.

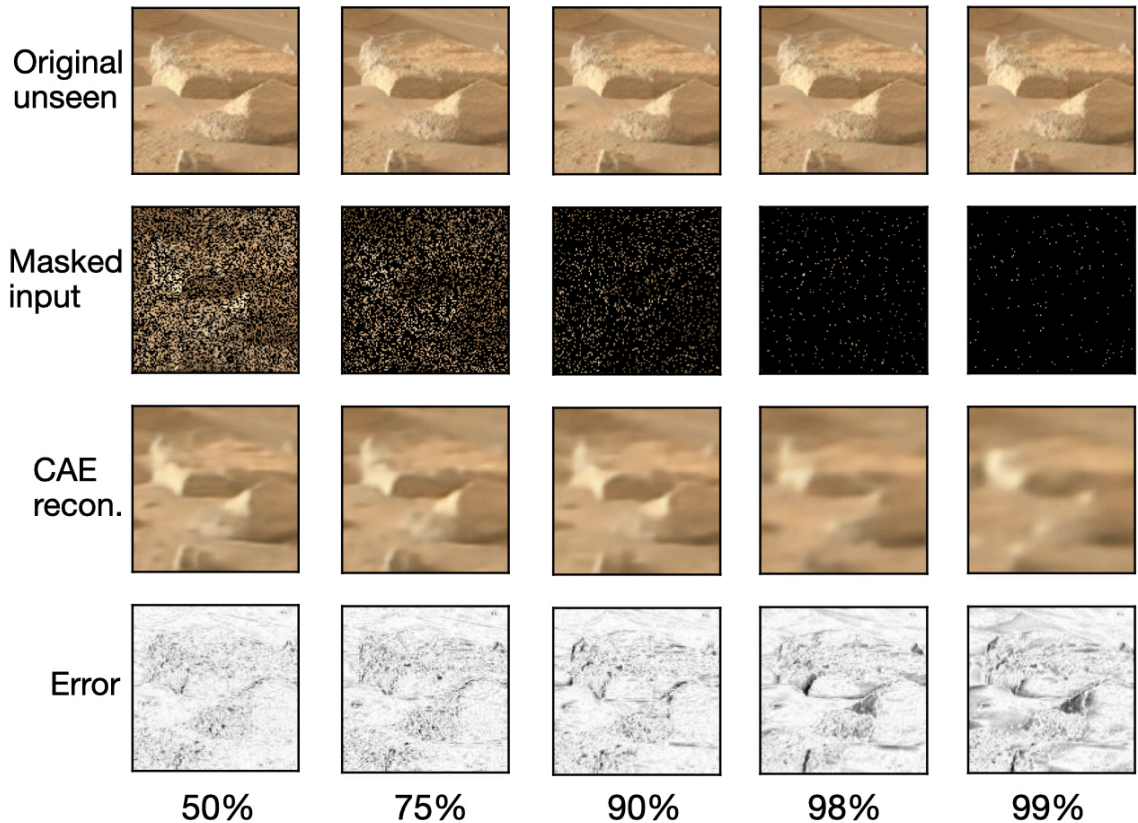

**Figure 3**: *Example outcomes for CAEs trained on destructive luminance noise of fixed areal coverage and then applied to unseen data. Top row: original image not used in training, 2nd row down: original image masked by destructive luminance noise, 3rd row: CAE reconstructed/generated image based on masked input only. Bottom row: error map formed from 3 channel pixel differences; darker features indicate larger errors. Noise masks are applied with 50%, 75%, 90%, 98%, and 99% coverage, left to right.*

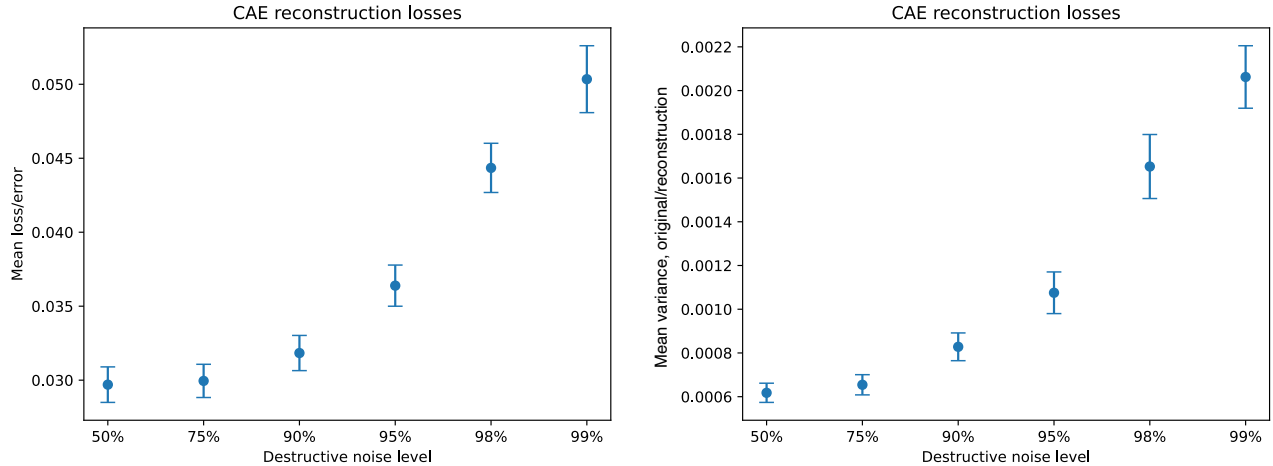

**Figure 4**: *(a) Left panel: Mean differences between original and reconstructed images across all pixels in 3 channels for an example CAE image reconstruction (c.f., Figure 3) as a function of luminance noise/masking level. Error bars correspond to 1-sigma errors on the mean. (b) Right panel: Mean variance in pixel-by-pixel differences between original and reconstructed images, across all images, as a function of luminance noise/masking level for the 6 models trained here.*

Using this metric the loss/error ranges from about 3% to 5% as the level of destructive noise increases from 50% to 99%. However, since these values reflect the level of mean (typical) differences, they are expected to remain small since the CAE, in the absence of informative input, will output values close to the statistical mean pixel values across all training images. A more informative metric may be the mean variance across all unseen inputs, shown in Figure 4b.

The variance metric (averaged over all input data) shows an increase of a factor of ~3.5 between 50% and 99% destructive noise levels. Consequently, both qualitative and quantitative assessments of reconstruction success indicate that the most significant degradation of the output occurs at around 90% noise and higher.

## 2.2.2 Training across luminance noise levels: a more flexible and more successful model: gt-CAE

Unlike many traditional analyses or data processing methods, deep learning can improve its efficacy when forced to generalize in training. Broadly speaking this is due to improvements in handling previously unseen features in new data after more exposure to feature complexity (e.g., Zeiler and Fergus 2013). To examine this capability a *single* CAE was trained (following the procedure used for the previous models, see above) on inputs with randomly chosen levels of pixel luminance noise/masking for each training subimage, with those areal coverage levels ranging from between 50% and 99%. The training and validation loss curves are plotted in Figure 5 for up to 40 training epochs, and indicate a cross-over in these curves at around 12 training epochs, where the model starts to overfit to the training data, and this is the training level used for subsequent analyses.

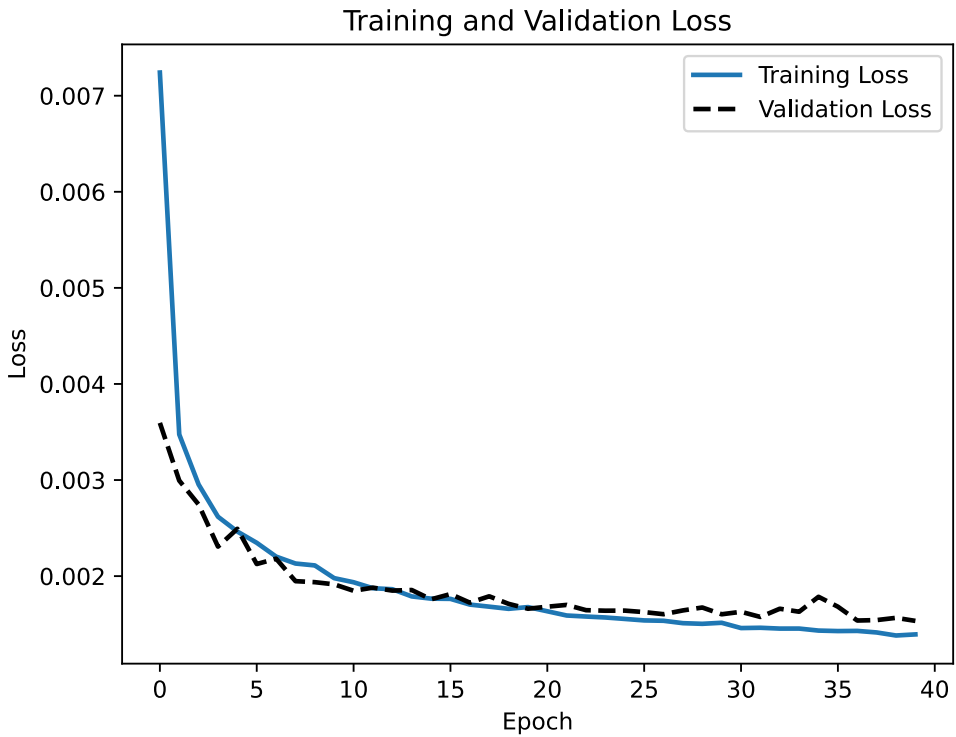

**Figure 5**: *Loss curves for single model, generally trained CAE (gt-CAE) training on all noise levels ranging from 50-99%.*

Figure 6 illustrates several reconstructions of previously unseen inputs at different noise levels by this more generally trained CAE (gt-CAE), as well as a comparison with the fixed-noise trained models. Performance of the more flexible gt-CAE appears excellent, and differences between it and the fixed-noise models are qualitatively and quantitatively minor and mostly in the finer details of how features are reconstructed. This indicates that a single, generally trained model can be successfully applied to all levels of noise/masking in this type of imaging data.

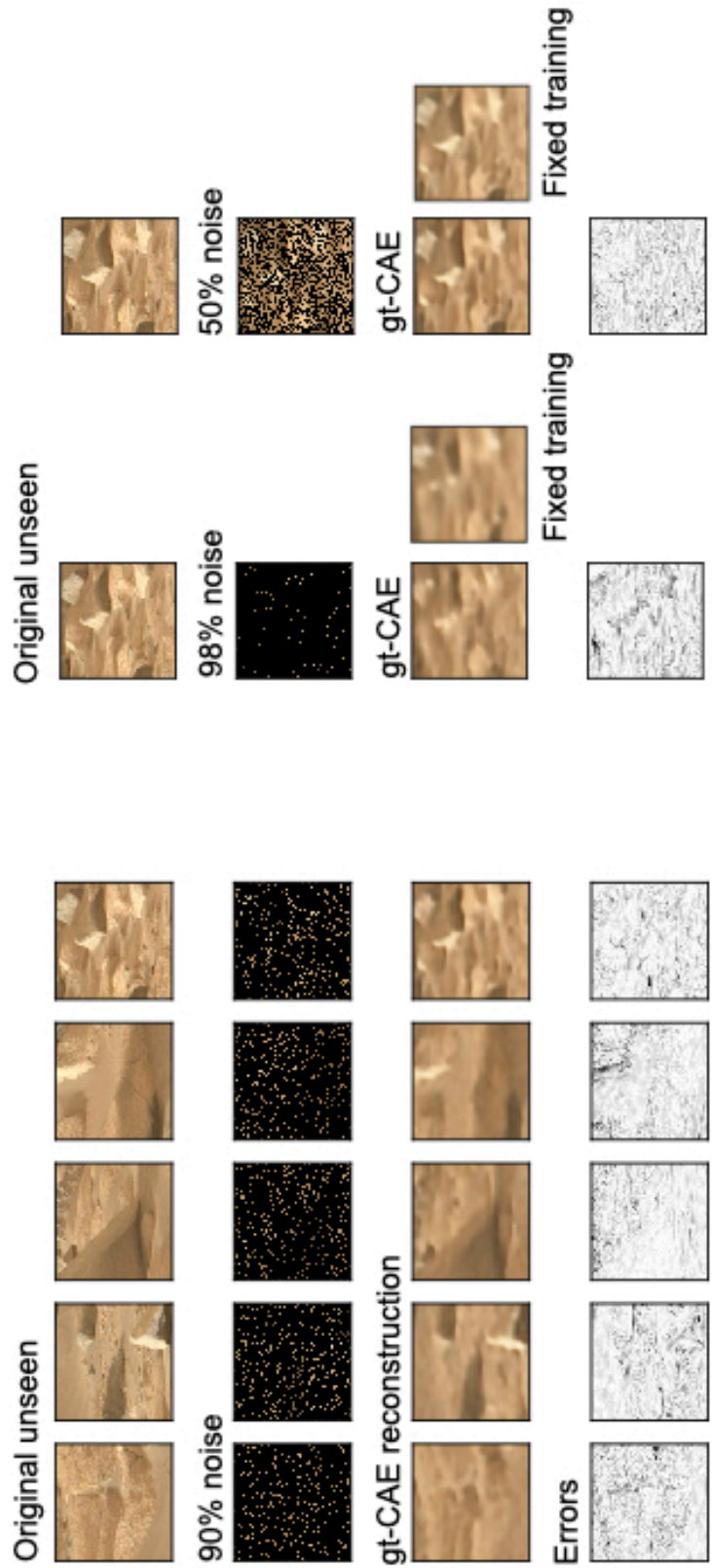

**Figure 6**: *Examples of gt-CAE reconstructions and cross-comparison with fixed noise level trained CAEs. **Left**: Results on attempted reconstructions of 5 unseen/validation subimages (originals at top) given as inputs to gt-CAE with 90% areal coverage destructive noise applied. Third row shows gt-CAE reconstructions, bottom row shows error maps calculated as previously. **Right**: gt-CAE reconstructions of unseen subimages with 98% and 50% noise applied, together with comparison to fixed noise level CAE results for training on 98% and 50% noise.*

### 2.2.3: Fully destructive luminance + color noise

While the destructive luminance noise/masking applied in the previous sections is challenging, inpainting across regions of zero-pixel values is a relatively constrained problem. A more common, and potentially even more challenging type of destructive noise may occur as an additive perturbation across pixel luminance/color values, e.g. due to instrument readout noise, real-time radiation-induced pixel variations, intrinsic pixel sensitivity variations, or data transmission errors. As a preliminary examination of CAE capabilities in handling this type of noise, a gt-CAE variant was trained following a similar procedure to Section 2.1.2, with training inputs subject to luminance + color noise applied to random pixels with total areal coverage between 10% and 100% of the images. Here, the luminance/color noise is modeled as an additive random value evenly distributed in the range [-x, x] across all channels (where x is set as a percentage of channel range), with a final pixel value in all channels clipped to [0,1]. I.e., an original pixel value may be randomly increased or decreased by the same amount in all 3 channels. Results from a 20 epoch gt-CAE trained this way with x=1.0 (100%) are presented in Figure 7 for a previously unseen input image.

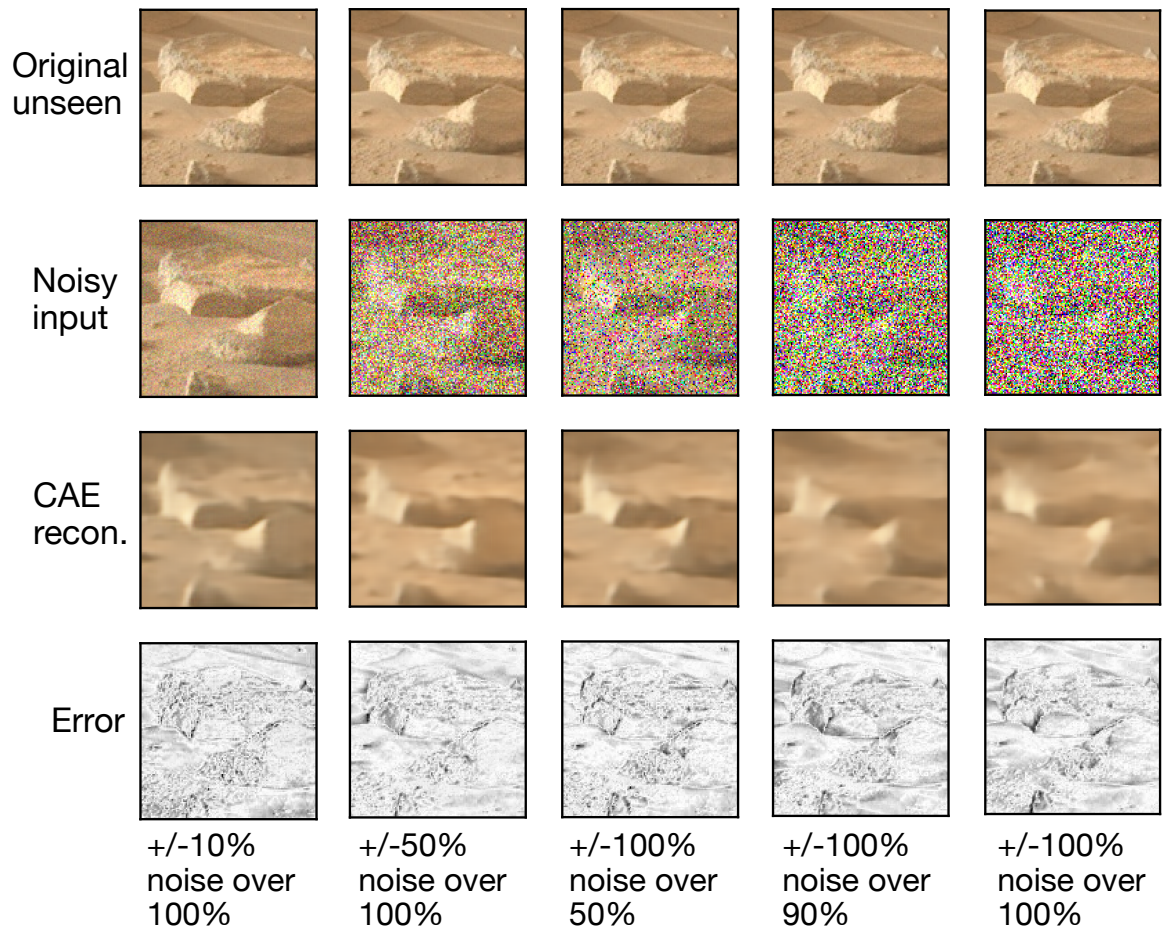

**Figure 7:** *Example outcomes for a gt-CAE trained on 10%-99% areal coverage fully destructive luminance + color noise (random +/- pixel perturbations) applied to unseen data. Top row: original image not used in training, 2nd row down: input image presented to trained gt-CAE including noise, 3rd row: gt-CAE reconstructed/generated image. Bottom row: error map formed from 3 channel pixel differences; darker features indicate larger errors. Noise is applied at levels of +/-10%, +/-50%, and +/-100% over random pixels with areal coverage indicated (from 50% to 100% random coverage). Pixel values are truncated to [0,1] in each channel.*

Although the model is trained on noise applied at levels of +/-100% of the original pixel range (i.e. random additive values within [-1,1]), the tests shown in Figure 7 apply a broader range of noise levels, including +/-10% (i.e. random additive values in the range [-0.1,0.1] and +/-50% [-0.5,0.5]. However, the reconstructions are remarkably robust to variations in the level of additive noise and the areal coverage of that noise. Even when 100% of pixels are subject to strong luminance + color noise, the gt-CAE model succeeds at generating a qualitatively recognizable reconstruction of the original, non-noisy, image, with an error map comparable to those in Figures 3 and 6.

While this success is significant, it does involve a relatively narrow and small training set. This type of denoising problem is also the basis of diffusion models (e.g., Ho et al. 2020), and for much larger, and more diverse datasets, diffusion models are likely to provide better performance and latent representations than a more traditional CAE approach.

**2.3 Out of bounds reconstruction: the sunflower test**

As a final test of the CAE and gt-CAE models used here, the responses to a fully anomalous input, or an "out of bounds" situation, are examined. An image with very different color scheme and different intrinsic scene features (a sunflower photograph, Figure 8) was given as input to: 1) the no noise/no mask trained CAE, 2) a fixed destructive luminance/masking noise trained CAE with noise at 98% (with the input also subjected to 98% areal coverage destructive random noise), and 3) the generalized luminance/masking noise trained gt-CAE (50-99% areal noise) presented with inputs subjected to 90% and 98% noise.

The direct reconstruction (no noise) CAE recovers most of the primary image features (brightness and contrast, separation of forms) but produces a very different color palette that matches the color palette of the Mastcam training set rather than the input sunflower data. At 98% noise the reconstructed images from both the fixed CAE and generalized gt-CAE shows severe spatial distortion, and the models are clearly projecting the input to a "best-guess" part of the latent space of the Mastcam training data. In other words, the autoencoder "sees what it knows" when presented with out of bounds data. It is noted that the gt-CAE does a slightly better qualitative job at 98% noise levels in terms of reconstructing shapes and patterning.

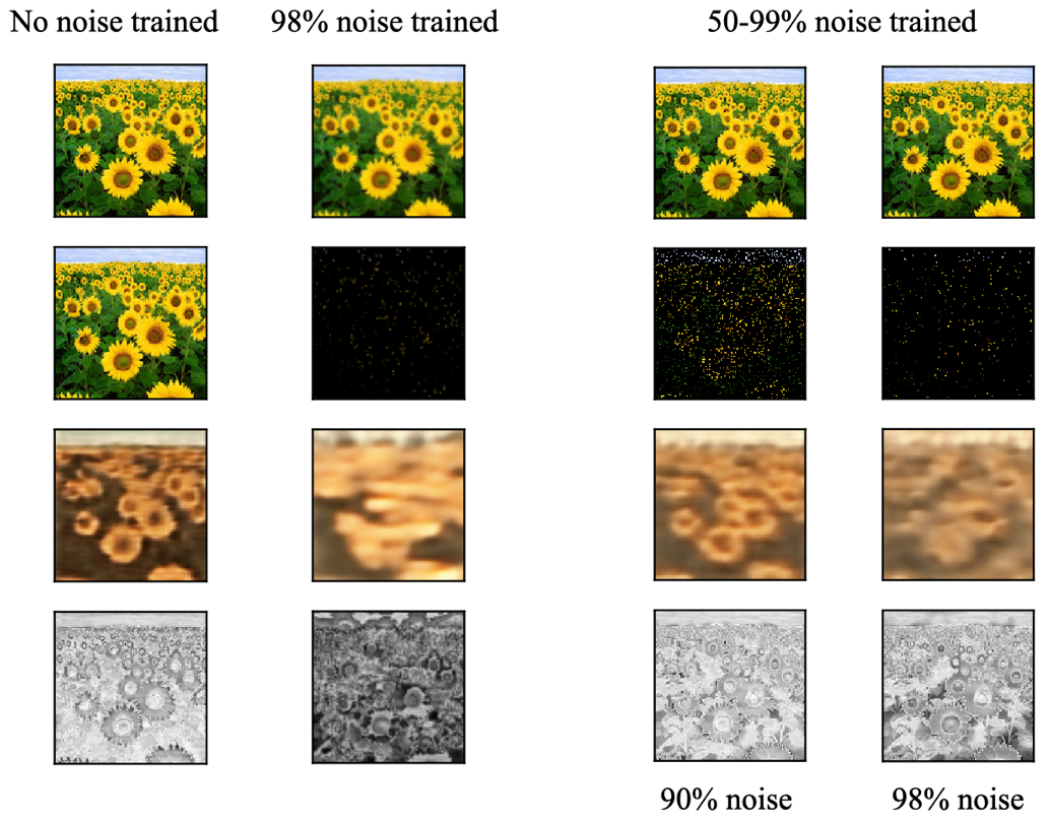

**Figure 8**: *Examples of the CAE and gt-CAE responses to out of bounds data. Left column: no noise CAE trained on the MastCam data used in this work, top row original unseen image, 2nd row down input version of image, 3rd row reconstruction, 4th row error map. Second column from left: input degraded by 98% noise and reconstructed by fixed 98% noise trained CAE. Third column from left: generalized gt-CAE presented with 90% noise degraded input. Rightmost column: generalized gt-CAE presented with 98% noise degraded input.*

Although extreme, this example indicates some points of caution in CAE applications. First, a lightweight CAE such as this, trained on a particular (narrow) class of images will reconstruct inputs accordingly, mapping to known parts of the latent space rather than discovering new classes. Second, if a CAE is to be used to discover anomalies or novelty in data it requires carefully constructed metrics to gauge those features, or incorporate additional learning schemes (e.g., classification layers). For example, a CAE could search for anomalies by masking data and comparing its reconstruction to the unmasked data, so that out of distribution data will be indicated by worse reconstruction errors. Or a CAE with classification layers could seek anomalies by examining similarity measures to classes/clusters in the latent space (see, for example, work in galaxy classification: Walmsley et al. 2022, Etsebeth et al. 2024.)

More generally, improvements to reconstruction behavior with out of bounds or out of distribution data can be made by pre-training models on very diverse data (e.g., Earth data might be used, or even the ImageNet visual database, see for instance Hendrycks et al 2019, Mantha et al. 2022. However, the best strategy is highly dependent on the goal. If, for example, a CAE or similar modeling approach is to be used for novelty or anomaly detection on Mars, then pre-training with diverse terrestrial data may be counterproductive, as novel, interesting features on Mars may be common in terrestrial data.

## 3.0 Discussion

### 3.1 Applications

The baseline CAE models presented here demonstrate an effective approach to reconstructing a class of planetary image data that might be impacted by destructive random luminance, and/or color noise, or deliberate masking. While these types of noise represent a specific form of image degradation, and other types will present different challenges (e.g. gaps), this work suggests that relatively lightweight models (with only ~$10^5$ internal parameters) such as CAEs have significant potential utility for exploration science. Such models could be more easily trained, or deployed after pre-training, on spacecraft or robotic platforms with limited computing resources.

Furthermore, if acceptable image reconstruction for a given purpose can be achieved with 90% destructive luminance noise (as shown here), the deliberate, pre-determined, random masking of an image (whether in software or hardware) could reduce data size by a similar amount. In effect, an instrument would image only what is needed for a CAE model to return useful reconstructions. Equally, if low illumination levels or obscurations (e.g., atmospheric conditions) drive some pixel values into the instrument noise, the CAE approach here could – if appropriately trained on expected data features - provide a "best possible" reconstruction of a scene to support a variety of exploration and scientific needs. In contrast to traditional lossy compression methods, this approach retains full accuracy and precision in stored data points (e.g., pixels), and reconstructs inputs by utilizing the full latent space of informative, contextual, data features.

A further possibility is also indicated by this work. In planetary and astrobiological science it is often an implicit assumption that a large corpus of imaging data is an important, if not indispensable, component of data acquisition for research needs. Images play a central role for human senses and for the ways in which we construct meaningful models of physical reality – either individually or as an ensemble species. A successful model of the world, in space, time, material composition and dynamical properties is used to navigate, assess, and predict with, because that model contains the essential, meaningful features of the world, and less of the uninformative qualities of the world. In this context, while there are perhaps an infinite number of procedural ways to use images to help form a model of the world, the latent space and representations as formed by a CAE can efficiently and uniquely reduce corpus data dimensionality and size, while preserving the most important feature information in a statistically robust fashion.

Consequently, there may be scenarios where the most valuable data returns from an instrument are the models and latent representations rather than a large dataset of traditional "raw" images (i.e., the model becomes the message). For example, consider a mobile surface mission that performs a wide-ranging environmental sweep of an unexplored landscape, with a goal of mapping the general properties of that landscape, and how those properties vary across some region. Power and/or communication constraints (e.g., in the outer solar system) may render a full areal coverage image return campaign impossible. In this scenario a CAE pretrained on analog landscapes [c.f. Kerner et al. 2029; Holland et al. 2022] could be used and fine-tuned in situ at low cost before returning its (much smaller) model, and a sampling of latent representations. In fact, if the ultimate scientific questions asked of the data involve the application of machine learning approaches, a latent space may have equal, if not greater, value than any corpus of raw imaging data.

That could be particularly true if a variational autoencoder is used. As described above, in a VAE an encoder defines an approximate posterior distribution $q(z|x)$, and outputs a set of parameters for specifying the conditional distribution of the latent representation $z$ for inputs $z$. The decoder trains to take the latent representation $z$ and (using its conditional distribution of $p(x|z)$) outputs the parameters of a conditional distribution of the input. In practice this also means that a trained variational autoencoder can be used to generate 'new' (fake) data with statistically valid features by sampling the probabilistic latent space, and can be interrogated to produce statistically robust answers to questions about the properties of phenomena of interest.

## 3.2 Foundation Models versus lightweight models

A new paradigm has emerged recently via the creation of Foundation Models (FMs) that are typically very large (multibillion, even trillion parameter) deep learning models trained on very large, broad data in an unsupervised or self-supervised fashion (e.g. applying the self-attention Transformer architecture to unlabeled data, Vaswani et al. 2023). A FM can learn features in data that are very broadly relevant and can then be subsequently fine-tuned to much more specialized tasks using far less new data and far less downstream compute time. In principle FMs could easily lead to the kinds of applications described here for CAEs with planetary imaging. For example, recent models resembling vastly expanded versions of gt-CAE have been developed for multispectral imaging tasks (e.g., SpectralGPT, Hong et al. 2024). However, the size of FM models (e.g., SpectralGPT has 600 million parameters), and both the initial compute investment, and later application can involve significant resources. Furthermore, the interpretability of end results produced from the highly abstracted latent representations in FMs is not as straightforward as with much simpler models (which can also still be challenging to understand).

Despite these considerations, the underlying architectures of FMs (e.g., Transformers, diffusion, or elaborations of CAEs such as U-Nets) can also be applied at smaller scales and can certainly outperform models such as simple CAEs. Exploring that middle-ground in the trade space of mission parameters will be important. There is also good reason to suppose that standalone, relatively specialized, lightweight ML applications will continue to be relevant for scientific exploration. Simpler, slower, resilient, flight-tested CPUs and onboard data storage and power budgets, create a challenging trade space where small differences in efficiency and efficacy can be critical.

## 3.3 Future directions

The work presented here is a simple proof-of-concept for relatively uncomplicated CAEs applied to data subject to destructive luminance and color noise, or deliberate masking/sparse-sampling. For further applications the details of the model's efficacy should be better quantified, e.g., pixel-by-pixel error maps have limited utility in determining the value of reconstructed data, since different features will have different importance. For example, it may be a priority to established local topography at certain physical scales (where accurately reconstructing larger features is key, but finer details are not necessary), or it may be a priority to retain accuracy in multichannel color features. Similarly, the architecture and hyperparameters of the CAEs used here could be further investigated to establish whether the present model is close to optimal in terms of return-on-compute, or in terms of generalizability to adjacent datasets (i.e., terrain imaging from different martian locations). That could be extended to examine what a minimal CAE would be, i.e., the smallest, least computationally demanding model to train or use that still accomplishes its task. A minimal CAE could inform decisions for designing onboard mission computing needs.

For the use case of deliberate sparse sampling/masking, it is not clear *a priori* that random sampling (i.e. as modeled here) is optimal for capturing essential features and ensuring the best reconstruction possible. Image or data sampling strategies belong to a very broad range of data science problems, and have been widely studied across many fields (e.g., Nyquist-Shannon sampling, see Unser 2000). One option that could be explored is a form of reinforcement learning (Szepesvári 2009) where metrics of success at image reconstruction (which would need to be defined) are fed back as input to a sampling model, and subsequently optimized for CAE use. In other words, the CAE itself learns what the best data sampling strategy is.

Finally, to test the idea of returning model weights and latent representations rather than traditional "raw" imaging data, work could be done to determine the efficacy of latent representation data for relevant scientific analyses. For example, if hyperspectral imaging data are transmitted, and subsequently used to evaluate mineralogical compositions and distributions, with an established analysis scheme (e.g. a variational CAE performing segment classification), what would be the impact on those end results if, instead, only latent representations were transmitted? To assess this impact, a comparison of scientific return would be needed between a traditional data approach and pipeline (with raw or minimally processed data returned, reduced, analyzed, and conclusions drawn), and the "model only" approach suggested here, where the corpus of data undergoes – in effect - intelligent compression that captures all of its informative features.